\documentclass[%
aps,
pra,
reprint,
nofootinbib,
amsmath,
amssymb,
longbibliography
]{revtex4-2}
\usepackage{
    physics,
    graphicx,
    multirow,
    mathtools, 
    placeins, 
    nicefrac, 
    hyperref 
}
\usepackage{adjustbox}
\usepackage[dvipsnames]{xcolor}
\hypersetup{
    colorlinks,
    linkcolor={blue!50!black},
    citecolor={blue!50!black},
    urlcolor={blue!80!black}
}

\usepackage[caption=false]{subfig} 
\usepackage[capitalize]{cleveref}

\DeclarePairedDelimiter\pbra{\langle\!\langle}{\rvert}
\DeclarePairedDelimiter\pket{\lvert}{\rangle\!\rangle}
\DeclarePairedDelimiterX\pbraket[2]{\langle\!\langle}{\rangle\!\rangle}{#1 \delimsize\vert #2}

\usepackage{enumitem}

\begin{document}
    
    
    \title{Resource-efficient Purification-based Quantum Error Mitigation}
    
    
    \newcommand{\addressOxMat}{\affiliation{Department of Materials, University of Oxford, Oxford, OX1 3PH, United Kingdom}}
    \newcommand{\addressOxStJohn}{\affiliation{St John's College, University of Oxford, Oxford, OX1 3JP, United Kingdom}}
    \newcommand{\addressQMT}{\affiliation{Quantum Motion Technologies Ltd, Nexus, Discovery Way, Leeds, West Yorkshire, LS2 3AA, United Kingdom}}
    \author{Zhenyu Cai}\email{cai.zhenyu.physics@gmail.com}\addressOxMat\addressOxStJohn\addressQMT
    
    \date{\today}
    
    \begin{abstract}
        To achieve the practical applications of near-term noisy quantum devices, low-cost ways to mitigate the noise damages in the devices are essential. In many applications, the noiseless state we want to prepare is often a pure state, which has recently inspired a range of purification-based quantum error mitigation proposals. The existing proposals either are limited to the suppressions of only the leading-order state preparation errors, or require a large number of long-range gates that might be challenging to implement depending on the qubit architecture. This article will provide an overview of the different purification-based quantum error mitigation schemes and propose a resource-efficient scheme that can correct state preparation errors up to any order while requiring only half of the qubits and less than half of the long-range gates compared to before.
    \end{abstract}
    
    \maketitle

    \section{Introduction}\label{sec:intro}
    With the recent rapid advance of quantum hardware, we naturally would seek to find applications for these near-term devices. However, given the level of noise present in the hardware and the limitation in the qubit number, it is still challenging to implement full quantum error correction on them. Hence, instead we turn to another range of techniques called \emph{quantum error mitigation (QEM)}, which in general try to recover the measurement statistics of the ideal observable by using additional circuit runs on the noisy machine. Techniques like this includes error extrapolation~\cite{liEfficientVariationalQuantum2017,temmeErrorMitigationShortDepth2017,endoPracticalQuantumError2018,caiMultiexponentialErrorExtrapolation2021}, quasi-probability method~\cite{temmeErrorMitigationShortDepth2017,endoPracticalQuantumError2018} and symmetry verification~\cite{mcardleErrorMitigatedDigitalQuantum2019, bonet-monroigLowcostErrorMitigation2018,caiQuantumErrorMitigation2021}. 
    
    More recently, using the fact that the ideal state we want to prepare in the experiments is often a pure state and the noise will corrupt it into some mixed state, a range of \emph{purification-based QEM} schemes are proposed to extract the pure state component of the noisy state we have. There are largely two types of schemes. One uses multiple copies of the noisy state~\cite{koczorExponentialErrorSuppression2020,hugginsVirtualDistillationQuantum2021}, which can suppress the state preparation noise exponentially with the increase of the number of copies. However, it would require long-range gates among the copies, whose noise are not suppressed by the purification. The other~\cite{obrienErrorMitigationVerified2021,huoDualstatePurificationPractical2021} tries to extract the pure state component by performing a projective measurement into the ideal pure state at the end, which does not require many gates beyond state preparation. However, due to gate noise we will get a `projection' into a noisy state instead and it can only remove the leading-order errors in the state preparation. In this article, we will provide an overview of these purification-based QEM schemes and their respective constraints, then we will see that the simple idea of combining these schemes can overcome the constraints outlined above.
    
    \section{Purification-based Quantum Error Mitigation}
    In this section, we will discuss the general idea of using purification for QEM, which is first discussed by Koczor~\cite{koczorExponentialErrorSuppression2020} and Huggins~\textit{et~al.}~\cite{hugginsVirtualDistillationQuantum2021}
    \subsection{State Purification}\label{sec:pure_state}
    We are given a circuit whose noiseless output is some pure state $\rho_0 =\ket{\psi_0}\bra{\psi_0}$, but in practice the circuit will output some mixed state $\rho$ due to the noise in the circuit. If the noise in the circuit is not so large such that the information of the ideal state $\rho_0$ is mostly erased, we would expect the dominant eigenvector of $\rho$ to be some state close to $\rho_0$. First let us assume the dominant eigenvector of $\rho$ is $\rho_0$ for simplicity while the more general case will be discussed later. The eigen-decomposition of the noisy state $\rho$ can now be partitioned into the ideal eigenstate $\rho_0$ and the noisy component $\rho_{\epsilon}$ which is a weighted sum of all the other eigenstates:
    \begin{align}\label{eqn:noisy_decomp}
        \rho = (1 - p_\epsilon)\rho_0 + p_\epsilon \rho_{\epsilon}.
    \end{align}
    Here $p_\epsilon$ is the infidelity of our original noisy state $\rho$. Since the eigenvectors of $\rho$ are orthogonal to each other, we have:
    \begin{align}\label{eqn:comp_ortho}
        \rho_0\rho_\epsilon = \rho_\epsilon\rho_0= 0.
    \end{align}
    The purified state is defined as:
    \begin{align*}
        \rho_{pur}^{(M)} = \frac{\rho^M}{\Tr(\rho^M)} = \frac{\left(1-p_\epsilon\right)^M\rho_0 + p_\epsilon^M \rho_\epsilon^M}{\left(1-p_\epsilon\right)^M + p_\epsilon^M \Tr(\rho_\epsilon^M)},
    \end{align*}
    where $M$ is the \emph{degree of purification}. Its fidelity against the ideal state $\rho_0$ is simply
    \begin{equation}
        \begin{split}\label{eqn:pur_fid}
            F^{(M)} = \Tr(\rho_0\rho_{pur}^{(M)}) &= \frac{ \left(1-p_\epsilon\right)^M}{\left(1-p_\epsilon\right)^M + p_\epsilon^M \Tr(\rho_\epsilon^M)} \\
            &\geq  \frac{\left(1-p_\epsilon\right)^M}{\left(1-p_\epsilon\right)^M + p_\epsilon^M}.
        \end{split}
    \end{equation}
    Hence, the infidelity of the $M$th-degree purified state is:
    \begin{equation}
        \begin{split}\label{eqn:pur_inf}
            P_\epsilon^{(M)} = 1 - F^{(M)} &\leq \frac{p_\epsilon^M}{\left(1-p_\epsilon\right)^M + p_\epsilon^M}\\
            & = p_\epsilon^M + \order{p_\epsilon^{M + 1}}
        \end{split}
    \end{equation}
    i.e. the infidelity has reduced from $p_\epsilon$ for the original noisy state $\rho$ to $p_\epsilon^M$ for the $M$th degree purified state $\rho_{pur}^{(M)}$, which is suppressed exponentially as the degree of purification $M$ increases if $p_\epsilon$ is small. Note that, even if $p_\epsilon$ is not very small, we will still have $P_\epsilon^{(M)} \leq p_\epsilon$ as long as $p_{\epsilon} \leq 0.5$. Hence, applying purification can move our input noisy state closer and closer to our ideal state $\rho_0$ with the increase of the degree of purification $M$. 
    
    More exactly, purification will move our input noisy state closer to its dominant eigenvector and so far we have assumed this dominant eigenvector to be our ideal state. Deviation from this assumption simply means that even with an infinite degree of purification, which returns the perfect dominant pure state, the lowest infidelity we can achieve is still bounded by the infidelity between the dominant pure state and our ideal pure state. This is called \emph{noise floor} in Ref.~\cite{hugginsVirtualDistillationQuantum2021} and \emph{coherent mismatch} in Ref.~\cite{koczorExponentialErrorSuppression2020}. Such coherent mismatch can be shown to be exponentially smaller than the incoherent error that we can suppress in practically relevant cases~\cite{koczorDominantEigenvectorNoisy2021}. Since in this article, we will mainly be comparing among different purification-based QEM schemes and the coherent mismatch will affect \emph{all} of them largely same way, we will not discuss the effect of coherent mismatch in this article.
    
    Instead of looking at $\rho^M$, which is the $M$th power of one single noisy state $\rho$, the above arguments are also applicable to the product of $M$ similar states $\prod_{m=1}^M \rho_m$. What we have discussed will work as long as all of these states have similar dominant eigenvectors, which would be true if they are prepared by similar unitary circuits (even with different noises)~\cite{koczorExponentialErrorSuppression2020,koczorDominantEigenvectorNoisy2021}. These could be the states prepared using the same circuit on different quantum registers (thus different noises in the circuits) or the dual state $\overline{\rho}$ that we will discuss later.

    \subsection{Purified Observable}\label{sec:pur_obs}
    Since the purified state $\rho_{pur}^{(M)}$ has a smaller infidelity against the ideal state $\rho_0$ compared to the original noisy state $\rho$, we would want to use $\rho_{pur}^{(M)}$ instead of $\rho$ for measuring the observable of interests $O$ in order to minimise errors. The expectation value of $O$ with $M$th degree purification is simply:
    \begin{align}\label{eqn:vir_dis}
        \Tr(O\rho_{pur}^{(M)}) = \frac{\Tr(O\rho^M)}{\Tr(\rho^M)}.
    \end{align}
    
    The bias of the expectation value of the purified observable in \cref{eqn:vir_dis} compared to the ideal expectation value $\Tr(O\rho_0)$ is indicated by the infidelity of the purified state $\rho_{pur}^{(M)}$ against the ideal state $\rho_0$ in \cref{eqn:pur_inf}. Hence, the bias of the purified expectation value should be suppressed exponentially with the increase of the purification degree $M$.
    
    The variance in the purified observable in \cref{eqn:vir_dis} is mostly related to the normalisation factor $\Tr(\rho^M)$, which indicates the strength of the signal of the purified state we want contained in the noisy state we start with~\cite{hugginsVirtualDistillationQuantum2021,koczorExponentialErrorSuppression2020}. The weaker the signal (small $\Tr(\rho^M)$), we would expect the larger the variance and thus the larger the sampling overhand. Therefore, the sampling overhead is also mostly determined by the input noisy state $\rho$ and the degree of purification $M$.
    
    For all of the QEM schemes that we will introduce later, they are essentially schemes for measuring $\Tr(O\rho^M)$ in \cref{eqn:vir_dis}. The denominator $\Tr(\rho^M)$ can be obtained using the same scheme with $O = I$ and then we can obtain the purified observable using \cref{eqn:vir_dis}. The purification process can only suppress noise in the state preparation of $\rho$. If there are additional circuit structures needed to measure $\Tr(O\rho^M)$, then the noise due to these circuit structures are not mitigated, which can increase our estimation bias and sampling overhead. Hence, one of our key goals is to find a purification scheme with the minimal additional gate cost for a given degree of purification. 
    
    \section{Different Purification Schemes}
    \subsection{Purification using Multiple Noisy Copies}\label{sec:multi_copy}
    \begin{figure*}[htbp]
        \centering
        \subfloat[Multi-copy Purification]{\includegraphics[width = 0.43\textwidth]{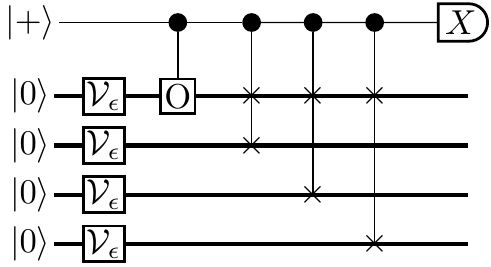}} \quad\quad\quad
        \subfloat[Multi-copy Purification with State Verification]{\includegraphics[width = 0.4\textwidth]{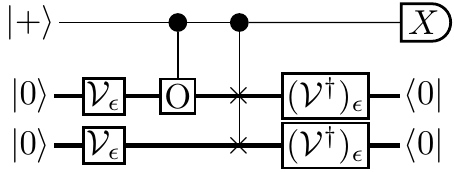}}\\
        \subfloat[Staggered Multi-copy Purification with Recycled Qubits]{\includegraphics[width = 0.7\textwidth]{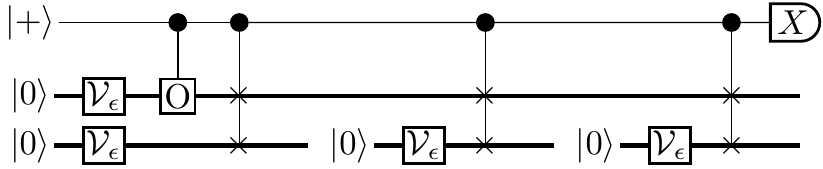}}
        \caption{The circuits for carrying out $4$th degree purification under different schemes. Each thicken line represents a whole quantum register. Hence, the $\ket{0}$ states are $\ket{0}$ of the whole quantum register instead of just one single qubit and the control-swaps are actually swapping the whole registers. Here we have illustrated circuits with their intermediate measurements performed using Hadamard test, but there are also circuit variants without needing any ancilla qubits as outlined in \cref{sec:prod_meas}.
        }
        \label{fig:vir_dist_circ}
    \end{figure*}
    \begin{figure}[htbp]
        \centering
        \includegraphics[width = 0.3\textwidth]{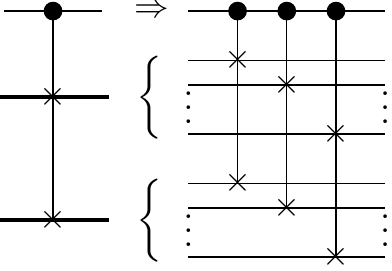}
        \caption{Decomposition of control-swaps between quantum registers into control-swaps between qubits.}
        \label{fig:cswap}
    \end{figure}
\begin{figure}[htbp]
        \centering
        \includegraphics[width = 0.3\textwidth]{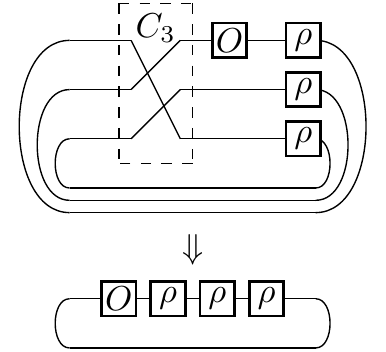}
        \caption{A diagrammatic proof for \cref{eqn:vir_dis_circ} with $3$ copies of the noisy state.}
        \label{fig:virdist}
    \end{figure}

    The way to measure $\Tr(O\rho^M)$ using $M$ copies of the noisy state $\rho$ was first introduced in Refs~\cite{koczorExponentialErrorSuppression2020, hugginsVirtualDistillationQuantum2021} by the name of \emph{error suppression by derangement} and \emph{virtual distillation}. In this article, to differentiate it from the other purification schemes, we will simply call it \emph{multi-copy purification}. The corresponding circuit is shown in \cref{fig:vir_dist_circ}~(a). Starting with $M$ copies of of the noisy state $\rho$, we will perform measurements of the observable $C_MO_1$ on the states, where $O_1$ is $O$ acting on copy $1$ and $C_M$ is the \emph{cyclic permutation operators} among the $M$ copies. The expectation value we obtained will then be:
    \begin{align}\label{eqn:vir_dis_circ}
        \Tr(C_MO_1\rho^{\otimes M})  = \Tr(O\rho^M).
    \end{align}
    A diagrammatic proof for the equation here is shown in \cref{fig:virdist}. 
    
    As seen from \cref{fig:vir_dist_circ} (a), we need to implement control-swaps among different copies, which can be decomposed into physical control-swaps acting on \emph{every} pairs of corresponding qubits among different copies as shown in \cref{fig:cswap}. In a typical $2$D qubit layout, the corresponding qubits in different copies would be far away from each other, thus these physical control-swaps would be long-range. For architectures that only allow short-range interaction, to perform each physical control-swap using local gates, we would need to first interlace different copies by swapping the qubits. Other schemes for the two-copy case that avoid using control-swaps for indirect measurements~\cite{hugginsVirtualDistillationQuantum2021} will still face the same connectivity issue and will require some other long-range gates. As discussed in \cref{sec:pur_obs}, these errors from the control-swaps and the swap gates used to overcome the connectivity issue are not suppressed in the purification scheme, thus it would be the main bottleneck for the bias and the sampling overhead of the purification scheme. Since the number of the additional long-range control-swaps needed increases with the number of copies $M$, the bias in the observable would not decrease indefinitely with the increase of $M$, there will be an optimal $M$ where the bias is the smallest. 
    
    \subsection{Purification using State Verification}\label{sec:state_ver}
    Now one must wonder can we perform purification without using additional copies of the quantum state. Refs.~\cite{obrienErrorMitigationVerified2021,huoDualstatePurificationPractical2021} has introduce ways to perform $2$nd degree purification via \emph{state verification}. Let us look at the denominator of the purified observable in \cref{eqn:vir_dis} for $M = 2$:
    \begin{align}\label{eqn:dist_ver}
        \Tr(\rho^2 O )  =\Tr(\rho O \rho).
    \end{align} 
    
    We see that it can be viewed as measuring the observable $\rho O$ on the state $\rho$. We cannot measure $\rho O$ directly. However, if we can perform measurements of $\rho$, then we should be able to measure $\rho O$ using Hadamard tests or non-destructive measurements of $O$ as outlined in \cref{sec:prod_meas}. To see how $\rho$ can be measured, let us first look at how $\rho$ is built. Let us suppose the perfect state $\rho_0$ is prepared using the perfect circuit $\mathcal{V}$ and the noisy state $\rho$ is prepared using the noisy version of the circuit $\mathcal{V}_{\epsilon}$:
    \begin{align*}
        \pket{\rho_0} &= \mathcal{V} \pket{0}\\
        \pket{\rho} &= \mathcal{V}_{\epsilon} \pket{0},
    \end{align*}
    where we have used the process matrix notation. Since we can implement projective measurements into the ideal state $\rho_0$ by applying the perfect inverse circuit $\mathcal{V}^\dagger$ and measure the projector into the state $\pket{0}$:
    \begin{align*}
        \pbra{\rho_0} &=  \pbra{0}\mathcal{V}^\dagger,
    \end{align*}
    we might think that we can perform measurement of $\rho$ by applying the inverse of the noisy channel $(\mathcal{V}_{\epsilon})^\dagger$ and measure the projector into the state $\pket{0}$. However, in practice, we cannot implement the inverse of the noisy circuit $(\mathcal{V}_{\epsilon})^\dagger$, we can only implement the noisy version of the inverse circuit $(\mathcal{V}^\dagger)_\epsilon$, which means that the measurement we perform at the end is actually:
    \begin{align*}
        \pbra{\overline{\rho}}  =  \pbra{0}(\mathcal{V}^\dagger)_\epsilon.
    \end{align*}
    Here the state $\overline{\rho}$ is simply the state obtained by apply the dual channel of the noisy inverse circuit:
    \begin{align*}
        \pket{\overline{\rho}}  =  ((\mathcal{V}^\dagger)_\epsilon)^\dagger \pket{0}
    \end{align*}
    and thus is named the the dual state of $\rho$ in Ref.~\cite{huoDualstatePurificationPractical2021}. Note that the dual channel $((\mathcal{V}^\dagger)_\epsilon)^\dagger$ is completely positive but might not be trace preserving (unless $(\mathcal{V}^\dagger)_\epsilon$ is unital). Therefore, the dual state $\overline{\rho}$ is positive semi-definite, but might not be normalised.
    
\begin{figure}[htbp]
        \centering
        \includegraphics[width = 0.3\textwidth]{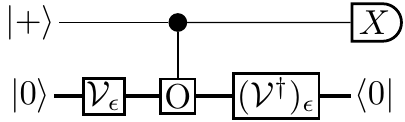}
        \caption{The circuit for measuring $\Tr(\overline{\rho} O\rho )$ in state verification. Here we have illustrated a circuit with the intermediate measurements performed using Hadamard test, but there are also circuit variants without needing any ancilla qubits as outlined in \cref{sec:prod_meas}.}
        \label{fig:state_ver}
    \end{figure}

    Hence, instead of measuring $\rho$ in \cref{eqn:dist_ver}, in practice we will be measuring the dual state $\overline{\rho}$, and thus the state-verified observable is:
    \begin{align}\label{eqn:sta_ver_O}
        \frac{\Tr(\overline{\rho} O\rho )}{\Tr(\overline{\rho}\rho)}
    \end{align}
    which is simply $2$nd degree purification but with one of the $\rho$ replaced with the dual state $\overline{\rho}$. The circuit for measuring $\Tr(\overline{\rho} O\rho )$ is shown in \cref{fig:state_ver}.
    
    Since the circuit $\mathcal{V}_\epsilon$ and $(\mathcal{V}^\dagger)_{\epsilon}$ are applying similar unitary gates but inverse, we would expect the dominant eigenvectors of the noisy state $\rho$ and the dual state $\overline{\rho}$ are very similar~\cite{obrienErrorMitigationVerified2021,koczorDominantEigenvectorNoisy2021,huoDualstatePurificationPractical2021} (the coherent mismatch is small). Hence, as discussed at the end of \cref{sec:pure_state}, we would expect the purification using  $\overline{\rho}\rho$ to perform similarly as the usual $2$nd degree purification using $\rho^2$. 
    
    Since state verification only involves one copy of the state, we do not need to implement any control-swap operations or operations for overcoming the connectivity issue like in the multi-copy approach. The only error in the state verification that is not accounted for is the intermediate measurement of $O$ (which also exists in the multi-copy approach). As discussed in Ref.~\cite{huoDualstatePurificationPractical2021}, since we can transform the observable into single-qubit observable and absorb the transformation into the state preparation circuit of $\rho$ and $\overline{\rho}$, the intermediate measurement of $O$ will only be a direct single-qubit measurement or a single-qubit measurement with the help of one single ancilla. Hence, the errors on the measurements can be mitigated using measurement error mitigation~\cite{maciejewskiMitigationReadoutNoise2020} while the noise on the ancilla can be removed by performing state tomography and post-processing purification~\cite{huoDualstatePurificationPractical2021}. In such a way, all of the noise in state verification can be suppressed in some way. However, state verification can only suppress the leading-order state preparation errors since it can only achieve $2$nd degree purification.
    
    \subsection{Other Extensions of the Purification Schemes}\label{sec:extensions}
    \subsubsection{Multi-copy Purification with Recycled Qubits}
    Ref.~\cite{czarnikQubitefficientExponentialSuppression2021} introduce a way to reduce the number of copies needed for multi-copy purification, whose circuit is shown in \cref{fig:vir_dist_circ} (c). It is done by rearranging the multi-copy circuit in \cref{fig:vir_dist_circ} (a) to stagger the preparation of different copies, which enable the reuse of one of the quantum registers to prepare multiple copies one by one. In this way, only two quantum registers are needed for all degrees of purification, one is used as the master copy, while the other is used to prepare the rest of the $M-1$ copies one by one. However, the full circuit depth in this case is $M-1$ times larger than the original circuit, which poses constraints for its practical use due to qubit lifetime. It is also more sensitive to the time needed for qubit initialisation. Since it is a rearrangement of the original multi-copy purification circuit, no reduction of the gate number is achieved. Hence, compared to multi-copy purification, it is mainly a scheme to trade circuit depth for a saving in the number of copies needed. 
    
    \subsubsection{Phase Estimation with State Verification}
    In Ref~\cite{obrienErrorMitigationVerified2021} where state verification is used, they did not directly perform measurements on the observable of interests $O$. Instead they perform phase estimation on the expectation value of the unitary evolution operator with the Hamiltonian $O$ (assuming $O$ is Hermitian):
    \begin{align*}
        U(t) = e^{-iOt}.
    \end{align*}
    which would ideally give rise to the noiseless signal $\Tr(U(t)\rho_0)$ when we probe at different $t$. By fitting curves to  the signal $\Tr(U(t)\rho_0)$, we can extract the information about $\Tr(O\rho_0)$. With noise present, we will have the noisy signal $\Tr(U(t)\rho)$ instead. By applying state verification, we can suppress the some of the noisy component and obtained the state-verified signal 
    $\Tr(\overline{\rho}U(t)\rho)$. By fitting to $\Tr(\overline{\rho}U(t)\rho)$, we can obtain the state-verified noisy estimate of the expectation value of $O$. Note that instead of explicitly sample for the normalisation factor $\Tr(\overline{\rho}\rho)$ in \cref{eqn:sta_ver_O} and perform the normalisation, we can simply normalised the fitted curve of the signal $\Tr(\overline{\rho}U(t)\rho)$ directly such that it takes the value $1$ when $t =  0$ ($U = I$).
    \subsubsection{Linear Combination of Purified States}
    It is worth noting that there are generalisations of purification schemes by looking at a linear combination of different degrees of purification~\cite{caiQuantumErrorMitigation2021,xiongQuantumErrorMitigation2021,yoshiokaGeneralizedQuantumSubspace2021}, in particular Ref.~\cite{caiQuantumErrorMitigation2021} also go beyond the discussion about permutation symmetry among noisy copies and went on to discuss general symmetries of the ideal state. 
    
    \section{Resource-Efficient Purification}
\begin{figure}[htbp]
        \centering
        \includegraphics[width = 0.3\textwidth]{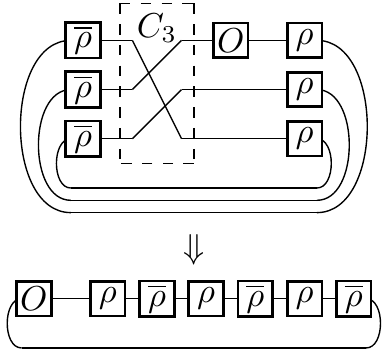}
        \caption{A diagrammatic proof for \cref{eqn:virdist_state_ver} with $3$ copies of the state.}
        \label{fig:virdist_statever}
    \end{figure}

    As discussed before, state verification can achieve $2$nd degree purification with much fewer additional gates than using $2$ noisy copies. We can extend the idea beyond $2$nd degree purification by applying state verification on top of multi-copy purification. The circuit is shown in \cref{fig:vir_dist_circ}~(b). We start with multiple copies of the noisy state $\rho^{\otimes M}$, then perform $C_MO$ measurement using the Hadamard test circuit like in multi-copy purification. After that, instead of discarding the states, we can further perform state verification on them, which gives:
    \begin{align}\label{eqn:virdist_state_ver}
        \Tr(\overline{\rho}^{\otimes M}C_MO_1\rho^{\otimes M}) = \Tr(O \left(\rho\overline{\rho}\right)^M)
    \end{align}
    This is shown diagrammatically in \cref{fig:virdist_statever}. With $O = I$, we can then measure the normalising constant $\Tr(\left(\rho\overline{\rho}\right)^M)$ and obtained the purified observable:
    \begin{align*}
        \frac{\Tr(O\left(\rho\overline{\rho}\right)^M)}{\Tr(\left(\rho\overline{\rho}\right)^M)}.
    \end{align*} 
    Hence, with the help of state verification, we can now perform $2M$th degree purification with only $M$ copies. 
    
    More importantly, the number of control-copy-swaps needed is reduced from $2M-1$ for multi-copy purification to $M-1$ for the combined scheme, a reduction by more than a half. Due to qubit number, connectivity and sampling overhead constraint, the number of copies used in purification cannot be too large in practice. For one of the most practically relevant cases in which we have two copies $M =2$, the combined scheme can achieve $4$th degree purification with only $1$ control-copy-swap, as opposed to in the $4$th degree multi-copy scheme where $3$ control-copy-swaps and $4$ copies are needed. The corresponding circuits are shown in \cref{fig:vir_dist_circ}. Hence, for a $4$th degree purification, the combined scheme reduces the number of qubits needed by half and the number of additional gates needed by $3$ times, which greatly enhance its practicality. It is worth noting that since we are using $2$ copies in this case, we can actually implement the measurement of $O$ and SWAP in place without using an ancilla as outlined in Ref.~\cite{hugginsVirtualDistillationQuantum2021}. If an ancilla is used, one should also consider removing the noise on the ancilla by applying the tomography purification scheme in Ref.~\cite{huoDualstatePurificationPractical2021}.
    
    Of course, it is also possible to apply state verification to only a subset of noisy copies and achieve odd-degree purification. In this way, our scheme can be further combined with the various extensions discussed in \cref{sec:extensions}.

    \section{Conclusion}
    In conclusion, for the existing purification-based QEM schemes, multi-copy purification would be limited by the number of qubits required and the additional gates needed to implement long-range control-swaps. On the other hand, state verification is only limited to $2$nd degree purification and thus can only suppress leading order errors. By applying state verification on top of multi-copy purification, we have a resource-efficient purification-based QEM scheme that can suppress errors to any orders while requiring only half of the qubits and less than half of the additional long-range gates compared to multi-copy purification. The saving is most pronounced in the practically relevant case of performing $4$th degree purification in which only half of the qubits and $1/3$ of the additional long-range gates are needed compared to before. Hence, the resource-efficient scheme should be one of the most practical choices for implementing purification-based QEM in real experiments.
    
    \section*{Acknowledgements}
    The author would like to thank Simon Benjamin and B\'alint Koczor for valuable discussions.
    
    The author is supported by the Junior Research Fellowship from St John’s College, Oxford and acknowledges support from Quantum Motion Technologies Ltd and the QCS Hub (EP/T001062/1).
    \appendix
    
    \section{Measuring Products of Operators}\label{sec:prod_meas}
    In this section, we will outline schemes for measuring the expectation value of the product of two operators.
    \subsection{Using Hadamard Test}
    This method will only work if one of the operators in the product is a unitary operator, which we will denote as $U$ and we would want to measure $SU$ where $S$ is some general Hermitian operator that we can directly measure. To perform the Hadamard test, we need to be able to perform control-$U$ gates, and of course direct measurement of $S$. 
    
\begin{figure}[htbp]
        \centering
        \includegraphics[width = 0.2\textwidth]{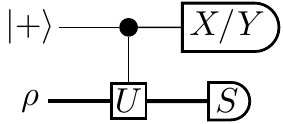}
        \caption{Hadamard Test Circuit. The expectation value of $X\otimes S$ will be $\Re(\Tr(SU\rho))$, while the expectation value of $Y\otimes S$ will be $\Im(\Tr(SU\rho))$.}
        \label{fig:Htest}
    \end{figure}

    For some incoming state $\rho$, we can measure the real part and the imaginary part of $\Tr(SU\rho)$ using the circuit in \cref{fig:Htest}. There the expectation value of the product of $X$ measurement on the ancilla and $S$ measurement on the data would give $\Re(\Tr(SU\rho))$. Replacing $X$ measurement on the ancilla with $Y$ measurement would give $\Im(\Tr(SU\rho))$. Note that this circuit measures the product $SU$ even if $S$ and $U$ does not commute, thus this is not equivalent to performing a non-destructive measurement of $U$ and then perform a measurement of $S$. i.e. the controlled-$U$ plus the ancilla preparation and measurement is not entirely equivalent to a non-destructive measurement of $U$ directly on the data qubit.

    \subsection{Using non-Destructive Measurement}
    The method here is introduced in Ref.~\cite{mitaraiMethodologyReplacingIndirect2019}. This method will only work if both operators in the product commute (which is trivial), or when one of the operators is an involution operator, i.e. it squares to the identity. Let us denote the involution operator as $G$, which we can measure non-destructively, and we want to measure $SG$ where $S$ is some general Hermitian operator that we can directly measure. By definition we have:
    \begin{align*}
        G^2 = I.
    \end{align*}
    Thus $G$ must have its eigenvalues squared to $1$, i.e. the eigenvalues are $\pm 1$. Hence, $G$ is Hermitian and unitary. The projection operator into the $\pm 1$ eigen-subspace of $G$ is simply:
    \begin{align*}
        \Pi_{\pm 1} = \frac{1 \pm G}{2}.
    \end{align*}
    The operator $S$ can be broken down into components that commute with $G$, denoted as $S_{+}$, and those anti-commute with $S$, denoted as $S_{-}$:
    \begin{align*}
        S = S_{+} + S_{-}.
    \end{align*}
    with
    \begin{align*}
        S_{+} = \frac{S + GSG}{2}\\
        S_{-} = \frac{S - GSG}{2}.
    \end{align*}
    For a given input state $\rho$, if we first measure $G$, the state will be projected into $\frac{\Pi_{\pm 1} \rho \Pi_{\pm 1}}{\Tr(\Pi_{\pm 1} \rho)}$ dependent on the measurement outcome. Now if we perform $S$ measurement on the state, we are essentially trying to perform a linear sum of measuring $S_+$ and $S_-$. Since $S_+$ commute with $G$, the measurement of $G$ will not affect the measurement result of $S_+$. If we measure the anti-commuting component $S_{-}$ on the resultant state after $G$ measurement, we would have:
    \begin{align*}
        \frac{\Tr(S_- \Pi_{\pm 1} \rho \Pi_{\pm 1})}{\Tr(\Pi_{\pm 1} \rho)} = \frac{\Tr( \Pi_{\pm 1} \Pi_{\mp 1} S_-\rho )}{\Tr(\Pi_{\pm 1} \rho)} =  0
    \end{align*}
    i.e. the $S_-$ measurement after $G$ measurement will produce expectation value $0$ regardless of the measurement outcome of $G$. 
    
    Hence, if we first measure $G$ and then measure $S$, the expectation value we get will only has contribution from the commuting part $S_+$. Thus, the product of the measurement results from the $G$ measurement and the $S$ measurement would be essentially measuring:
    \begin{align*}
        S_+G = \frac{SG + GS}{2}.
    \end{align*}
    This is equivalent to:
    \begin{align}\label{eqn:sym_meas}
        \sum_{\lambda = \pm 1} \lambda \Tr(S\Pi_{\lambda}\rho\Pi_{\lambda}) = \Tr(\frac{SG + GS}{2} \rho).
    \end{align}
    Similarly, if  we apply $R_{\pm} = e^{\pm iG\pi/4} = \frac{I \pm i G}{2}$ gate instead of measuring $G$ non-destructively, then we can combine the results of apply $R+$ and $R_-$ respectively and obtain:
    \begin{align}\label{eqn:anti_sym_meas}
        \sum_{\lambda = \pm 1} \lambda \Tr(SR_{\lambda}\rho R_{\lambda}) = i\Tr(\frac{SG - GS}{2} \rho).
    \end{align}
    We can combine \cref{eqn:sym_meas} and \cref{eqn:anti_sym_meas} to obtain $\Tr(SG\rho)$.
    
    For the application to state verification, it is enough to simply measure:
    \begin{align*}
        \frac{\Tr(O\left(\rho\overline{\rho} +\overline{\rho}\rho\right))}{2\Tr(\overline{\rho}\rho)} = \frac{\Tr(\frac{\overline{\rho}O +O\overline{\rho}}{2} \rho)}{\Tr(\overline{\rho}\rho)}
    \end{align*}
    which has the same purification degree and thus the same error suppression power as \cref{eqn:sta_ver_O}. Hence, the numerator $\Tr(\frac{\overline{\rho}O +O\overline{\rho}}{2} \rho)$ can simply be measured using non-destructive measurement as outlined in \cref{eqn:sym_meas} with $G = O$ and $S = \overline{\rho}$. The denominator can be measured directly with $O = I$.

%

\end{document}